\documentclass[11pt]{article}
    \usepackage{amsmath,amsxtra,amssymb, amsfonts, bbm} 
    \usepackage{color}

 \def\be{\begin{eqnarray}}
\def\ee{\end{eqnarray}}
\def\bee{\begin{eqnarray*}}
\def\eee{\end{eqnarray*}}

\def\pmx{\begin{pmatrix}}
\def\emx{\end{pmatrix}}
\def\bsq{\begin{subequations}}
\def\esq{\end{subequations}}

 \newcommand{\proj}[1]{ | #1 \kb  #1|}
 \def\ds{\displaystyle}

\def\ot{\otimes}
\def\tr{{\rm Tr} \, }
\def\trp{{\rm Tr} }
\def\bra{\langle}
\def\ket{\rangle}
\def\kb{ \ket \bra }
\def\raw{\rightarrow}

\def\half{\tfrac{1}{2}}
\def\wtd{\widetilde}
\def\nn{\nonumber}
\def\hil{{\mathcal H}}

\def\cG{{\cal G}}

  \title{Remarks\thanks{It 
  is noteworthy that Kim's note  arXiv:1210.5190  was posted very
  close to the 40th anniversary of the proof of SSA which was completed
  in October, 1972. } $~$ on Kim's  \\ Strong Subadditivity  Matrix Inequality: \\ 
  Extensions and Equality Conditions}
  
  \author{Mary Beth Ruskai\thanks{mbruskai@gmail.com} \\  
 Institute for Quantum Computing,  University of Waterloo \\ Waterloo, Ontario, Canada}  
  
  \date{\today}
  
\begin{document}

\maketitle

\begin{abstract}
We describe recent work of Kim in arXiv:1210.5190 to show that operator
convex functions associated with quasi-entropies can be used to prove a
large class of new matrix inequalities in the tri-partite and bi-partite setting by taking a
judiciously chosen partial trace over all but one of the spaces.  We 
 give some additional
examples in both  settings.
 Furthermore, we observe that the equality conditions for all the 
 new inequalities are essentially the
same as those for strong subadditivity.  
\end{abstract}


\tableofcontents


\section{Background}

In a recent paper Kim \cite{kim} showed that for operators on a tensor product
space $\hil_A \ot \hil_B \ot \hil_C $ one can obtain an interesting
new matrix inequality on  one space $\hil_C$ by taking the partial trace  $\trp_{AB}$
over a quantity for which the full $\trp_{ABC}$ would yield
 strong subadditivity (SSA) of von Neumann entropy.
A key ingredient is the inclusion of an additional operator  $K$ used in
the Wigner-Yanase-Dyson (WYD) skew information \cite{WY,Lb}, and then choosing a special
form for $K$.     As observed at the end of \cite{kim}, Kim's protocol can
be applied to a large class of operator convex functions, including those 
associated with the WYD skew information, to produce additional matrix
inequalities.   Although Kim used the recent elegant approach of Effros \cite{Ef},
earlier work, going back to Petz \cite{Pz1},   based on Araki's relative modular
operator \cite{Ak} will suffice.  
We remark at the end on the different approaches 
which lead to well-known properties   used below.

Kim's result seems remarkable in view of certain well known facts.
 The  concavity  
of the vonNeumann  entropy $S(\rho) = - \tr \rho \log \rho$ is an easy 
consequence of the much stronger operator convexity of $g(x) = x \log x$.
(See, e.g.,  \cite{Davis} and references therein.) 
However, it is also known that the  operator
$h(\rho, \gamma) =  \sqrt{\rho} \big(  \log \rho -\log \gamma \big) \sqrt{\rho} $
 is {\em not} jointly operator  convex, although one has separate operator
 convexity in the sense that the first term is operator convex in
 $\rho$ and the second in $\gamma$.  A trace is needed to obtain the joint convexity
of relative entropy $H(\rho, \gamma) = \tr \rho (\log \rho -\log \gamma)$.  
Thus, it is surprising that one can prove an operator inequality whose
trace would yield something which is an immediate corollary of
the joint convexity of relative entropy.  It should perhaps be emphasized
that this does not provide a new proof of SSA; rather, the new inequality
emerges from the proof of a mild strengthening of SSA by the inclusion
of an additional operator $K$ as in the WYD setting (from  which
the  usual relative entropy can be obtained as a
limit) and then making
a judicious special choice for $K$ in the tripartite setting.

As he observed, Kim's approach can be used with other convex operator
functions; some examples are worked out below.   
In addition, we show that in almost all cases, the equality conditions are identical to those in \cite{HJPW} 
for strong subadditivity.

\section{General theory}

\subsection{Some basics}

Let $\cG$ denote the class of operator convex functions $g(x) $ on $(0,\infty)$  with $g(1) = 1$.
For any $g \in \cG$, one can
define a generalized relative entropy \cite{LesR}, also known as an $f$-divergence \cite{ AN,HMPB} or
``quasi-entropy'' \cite{OP,Pz1} as
\be    \label{Hdef}
     H_ g(K, P, Q) \equiv \tr K^* g(L_P R_Q^{-1} ) R_Q(K) = \tr K^* g(L_P R_Q^{-1} ) K Q
\ee
where $P,Q > 0 $ are  positive definite matrices and $L_P (X) = PX$ and $ R_Q(X) = XR$
denote left and right multiplication respectively.   
It is by now well-known that the map $(P,Q) \mapsto     H_g(K, P, Q)$ is  jointly
convex in $P,Q$ for any pair of positive definite $P,Q$ and any fixed $K$.   
For $g(x) = - \log x$, 
\be   \label{relent}
    H_{- \log x}(K, P, Q)  = \tr  \Big(  K^* K  Q  \log Q -   K Q K^* \log P \Big)
\ee
which reduces to the usual relative entropy when $K = I$.

Whenever $g \in \cG$, then  $\wtd{g} \equiv x g(x^{-1})$ is also in $\cG$ and
 \be
     H_{\wtd{g}}(K^*, P, Q)  =   H_g(K,Q,P)  .
 \ee
In the case $g(x) = - \log x$ above, 
$\wtd{g} = x \log x $  and 
\be   \label{relentrev}
    H_{x \log x} (K, P, Q)  = \tr  \big( K K^* P  \log P - K^* P K \log Q \big)
\ee
 Because $K \neq I$ is important in what follows, we observe that
   for any function $f$  
\begin{align*}    
 \tr K^*  f( L_P)  K Q  & =    \tr K^* f(P) K Q  \\
 \intertext{but}
   \tr K^* f(R_Q)  K Q  & =  \tr K^* K  f(Q)   Q 
   \end{align*}
   which leads to the ``sandwiched'' expressions     $K Q K^*$ in  \eqref{relent}
    and  $K^* P K $  in \eqref{relentrev}.   This is inevitable
  unless $K$ happens to commute with $Q$ and/or $P$.

It is also well-known that  joint convexity  of  $H_g(P,Q)$ for fixed $K$ implies monotonicity under partial traces in the
following sense
\be  \label{monoa}
  H_g( I_A \ot K_{BC}  , P_{ABC}, Q_{ABC}) 
  & \geq &    H_g( K_{BC}  , P_{BC}, Q_{BC})        
\ee
and that one can not replace $I_A \ot K_{BC}$ by a general $K_{ABC}$.  The best
one can do is the minor generalization $V_A  \ot K_{BC}$ where $V_A$ is unitary \cite{JR}.

\subsection{Main result} 

 Using the definition \eqref{Hdef}, we can rewrite \eqref{monoa} as
 \be   \label{mono}
     \lefteqn{ \trp_{ABC} \, I_A \ot K_{BC}^*   \, g\big(L_{P_{ABC} } R_{Q_{ABC}}^{-1}  \big) K_{BC} \, Q_{ABC} }  \qquad \qquad  \nn \\
        &   \geq &  
          \trp_{BC} \,  K_{BC}^*   \, g\big(L_{P_{BC} } R_{Q_{BC}}^{-1}  \big) K_{BC}  \, Q_{BC} \\   \nn
          & = &  \trp_{ABC}  \,  K_{BC}^*   \, g\big(L_{P_{BC} } R_{Q_{BC}}^{-1}  \big) K_{BC}  \, Q_{ABC} 
\ee 
 where the equality follows from   $\trp_A  \, Q_{ABC} = Q_{BC}$ since there is  no
 explicit dependence on $\hil_A$ in the rest of the expression.    We also
 suppress tensor products with the identity so that, e.g.,  it is understood that  $P_{AB} $ means $P_{AB} \ot I_C $, etc.
Next, consider the special case $P_{ABC} = P_{AB} \ot I_C $ to get 
\be   \label{monob}
   \lefteqn{ \trp_{ABC} \,  I_A \ot K_{BC}^*   \, g\big(L_{P_{AB} } R_{Q_{ABC}}^{-1}  \big) K_{BC} \, Q_{ABC} }  \qquad \qquad  \nn \\
        &   \geq &  
         \trp_{ABC} \,  K_{BC}^*   \, g\big(L_{P_{B} } R_{Q_{BC}}^{-1}  \big) K_{BC}  \, Q_{ABC} 
\ee
 and choose $K_{BC} = I_B \ot K_C $. 
 Then $K$ commutes with $P_{ABC}  = P_{AB} \ot I_C $  so 
   that \eqref{monob} can be rewritten as
   \be   \label{monoc}
        \lefteqn{ \trp_{ABC} \,  I_{AB} \ot  K_C^* K_C    \, g\big(L_{P_{AB} } R_{Q_{ABC}}^{-1}  \big)   \, Q_{ABC} }  \qquad \qquad  \nn \\
        &   \geq &  
         \trp_{ABC} \,  I_{AB} \ot  K_C^* K_C       \, g\big(L_{P_{B} } R_{Q_{BC}}^{-1}  \big)  \, Q_{ABC} 
   \ee
  Furthermore, we can choose $K_C = \proj{\phi_C} $ to be a rank one projection so that  \eqref{monoc}
   becomes
 \be    \label{monod}
       \bra \phi_C ,   \trp_{AB } \Big[  \, g\big(L_{P_{AB} } R_{Q_{ABC}}^{-1}  \big)  -  g \big(L_{P_{B} } R_{Q_{BC}}^{-1}  \big) \Big]  \, Q_{ABC}   ~  \phi_C\ket   ~ \geq ~ 0
        \ee
where $| \phi_C \ket $ is an arbitrary vector in $\hil_C$.  This implies that the operator
   \be    \label{final}
    \lefteqn{      \trp_{AB } \Big[  \, g\big(L_{P_{AB} } R_{Q_{ABC}}^{-1}  \big)  -  g \big(L_{P_{B} } R_{Q_{BC}}^{-1}  \big) \Big]  \, Q_{ABC} } 
       \qquad   \qquad  \\  \nn
     & = &     \trp_{AB } \,   g\big(L_{P_{AB} } R_{Q_{ABC}}^{-1}  \big)  \, Q_{ABC} -  \trp_B  \, g \big(L_{P_{B} } R_{Q_{BC}}^{-1}  \big)  \, Q_{BC}  ~ \geq ~ 0
                 \ee
                 is positive semi-definite on $\hil_C$.
                 
                 We can choose $\hil_B$ to be one dimensional to obtain bipartite formulas, e.g., 
   \be    \label{final-bi}
         \trp_{A } \,   \Big[  \, g\big(L_{P_{AC}}^{-1}   R_{Q_{A}}   \big)  -   g \big(L_{P_{C}}^{-1}   \big) \Big]  \,  P_{AC} ~ \geq ~ 0
                 \ee
which is also useful.   However, we  presented the  development in
the tripartite situation because the most important application is to SSA.

 \subsection{Adjoint form}
                 
                 If one replaces $g(x)$ by  $\wtd{g} = x g(x^{-1})$  and  interchanges $P \leftrightarrow Q$, one obtains 
                   \be    \label{final-rev}
         \trp_{AB } \, P_{ABC}  \Big[  \, g\big(L_{P_{ABC}}^{-1}   R_{Q_{AB}}   \big)  -   g \big(L_{P_{BC}}^{-1}  R_{Q_{B}}  \big) \Big]  \,  ~ \geq ~ 0
                 \ee
     in which the LHS is formally the
                 adjoint of that in \eqref{final}; however, since any  positive semi-definite operator is
                 self-adjoint, \eqref{final-rev} is equivalent to \eqref{final}.
  For additional insight into why this is so, recall the basic property that $(WX)^* = X^* W^* $ reverses the order and hence,
   reverses left and right multiplication by  self-adjoint operators.   Thus
  \bee
    \big[ g(L_P R_Q^{-1})R_Q (X) \big]^* & = &    g(R_P L_Q^{-1}) L_Q (X^*)   \\
              & = &      L_Q R_P^{-1} g\big[ \big(L_Q R_P^{-1} \big)^{-1} \big] R_P(X^*)  \\
              & = &   \wtd{g} (L_Q R_P^{-1} ) R_P (X^*)
  \eee 
For $X = I_{AB} \ot \proj{\phi_C} = X^*$ the equivalence of \eqref{final} and \eqref{final-rev} is then clear.

                 
                 \section{Specific inequalities}
                 
                 \subsection{Subadditive type}   \label{sub:SSA}
                 
                 The choice $Q_{ABC} = \rho_{ABC},  P_{AB} = \rho_{AB} $ and $g(x) = - \log x $ in \eqref{final}
                 gives a result reminiscent  of SSA, i.e.,
               \bsq  \label{SSApos} \be   \label{SSAposA}
                    \trp_{AB}  \,  \big[ \log \rho_{ABC} - \log \rho_{AB} - \log \rho_{BC} + \log \rho_B \big]  \,  \rho_{ABC}~ \geq ~ 0
                 \ee
                  as an operator inequality on $\hil_C$.   Using, instead,  $\wtd{g}(x) = x \log x  = x g(x^{-1}) $  and \eqref{final-rev}
                  with the  choices $P_{ABC} = \rho_{ABC}, ~ Q_{AB} = \rho_{AB} $ gives  the result in the form written by 
                  Kim in \cite{kim}, i.e., 
                   \be  \label{SSAposB}
                      \trp_{AB} \,  \rho_{ABC} \big[ \log \rho_{ABC}- \log \rho_{AB}  +  \log \rho_B  - \log_{BC} \big]   ~ \geq ~ 0
                    \ee   \esq
                 which is formally the adjoint of \eqref{SSAposA}.   However, as remarked above, these are equivalent since
                 a positive semi-definite operator on $\hil_C$ is necessarily self-adjoint.
                If one uses  $\wtd{g} = x \log x $ without the exchange $P \leftrightarrow Q$, i.e, with the choice 
                $Q_{ABC} = \rho_{ABC},  P_{AB} = \rho_{AB} $,  
                     one gets                      
                      \be   \label{SSArev} 
                        \tr_{AB} \, \rho_{AB} \big[ \log \rho_{AB} - \log \rho_{ABC} - \log \rho_B + \log \rho_{BC}  \big] ~ \geq ~ 0
                     \ee
   in which the simple replacement of $\rho_{ABC} $ by $\rho_{AB}$ on the left appears to 
   reverse the usual form of SSA.   Note, however, that taking $\trp_C$ in \eqref{SSArev} does {\em not}
   yield SSA!
                 
                 When $\hil_{B} $ is one dimensional, \eqref{SSAposB} becomes 
                            \be            
                            \trp_A    \,  \rho_{AC}  \,    \big[ \log \rho_{AC} - \log \rho_A - \log \rho_C \big]    ~ \geq ~ 0   
                \ee     
               which is an operator version of ordinary subadditivity.
                   
                  \subsection{Relative entropy}  \label{sub:MPT}
                  
                   The more general choice  $Q_{AB} = \gamma_{AB} $ changes \eqref{SSArev} to
                    \be  \label{MPTineq}
                    \trp_{AB}  \,  \rho_{ABC} \big[ \log \rho_{ABC} - \log \gamma_{AB} - \log \rho_{BC} +  \log \gamma_B  \big]  ~  \geq ~ 0   
                 \ee
                 which is  naturally associated with the monotonicity of relative entropy under partial traces.
                 When  $\hil_B$ is one dimensional, this becomes
                 \be
                        \trp_{A}  \,  \rho_{AC} \big[ \log \rho_{AC} - \log \gamma_{A} - \log \rho_{C}  \big]  ~  \geq ~ 0   
               \ee
                  and for $\gamma_A = \tfrac{1}{d_A} I_A $ 
                  \be
                       -  \trp_{A}  \,  \rho_{AC} \log \rho_{AC} +  \rho_C \log \rho_{C}  \leq   (\log d_A )  \,  \rho_C  
                                          \ee
              which gives an upper bound on an operator version of conditional information, although this can
              be negative in the quantum setting.  
              
               Choosing $Q_{ABC} =  Q_{AB} \ot I_C $ in \eqref{final-rev} is essential
              to ensure that it commutes with $I_{AB} \ot K_C $.     This precludes a proof of the full-fledged 
              operator analogue of monotonicity of relative entropy  by this method since 
                           \be
                \trp_{A}  \,  \rho_{AC} \big[ \log \rho_{AC} - \log \gamma_{AC} - \log \rho_{C}  +  \log \gamma_C  \big]  
            \ee
           is {\em not} even Hermitian and, hence,  can
            {\em not} be positive semi-definite.
                   
          
          \subsection{WYD inequalities} \label{sub:WYD}
                      
                    The functions $g(x) =  \frac{1}{t(1-t)}(1 - x^t)$ and $\wtd{g}(x) =  \frac{1}{t(1-t)}  (x - x^{1-t})$   generate the WYD
                    skew information, for which $H_g(K,P,P) \geq 0 $ and $H_g(K,P,Q)  $ is jointly 
                     operator convex in  $P,Q $ in the maximal range $[-1,2] $ as observed
                     implicitly\footnote{Although first Lieb \cite{Lb} and then Ando \cite{ando}
                    obtained the key convexity result for the WYD entropy with $t \in (0,1)$, the seemingly innocuous omission of the
                    obvious linear term precludes writing their results in the general framework used here.  Ando 
                    also showed that the concavity
                    of Lieb's expression changes to convexity for $t \in (1,2]$.   \\} in \cite{ando}
                    and explicitly by  Hasegawa \cite{Has}.  (See also \cite{JR}.) 
                       Since, as is well known,  $\ds{\lim_{p \raw 1}  g(x) = - \log x}$ and
                  $\ds{\lim_{p \raw 0}  \,  \wtd{g}(x)= x \log x }$, one can also recover the results of Section~\ref{sub:SSA}.                   
             
                    Using  $g(x) $ in \eqref{monod} with  $Q_{ABC} = \rho_{ABC},$
                    and $P_{AB} = \gamma_{AB} $  gives the inequalities
                    \be     
                        \frac{1}{t(1-t)} \Big[ \trp_{AB} \, \rho_{ABC} - \trp_{AB} \,  \rho_{ABC}^{1-t}  \, \gamma_{AB}^t  \, - 
                            \trp_B \, \rho_{BC} - \trp_B \,   \rho_{BC}^{1-t}  \ \, \gamma_{B}^t \Big]  ~ \geq  ~ 0     \quad
                                            \ee
                 for any $ t \in [-1,2]$.   Since   $ \trp_{AB} \, \rho_{ABC}  =  \trp_{B} \, \rho_{BC} = \rho_C  $, this becomes
                                    \be      \label{WYD}
                        \frac{1}{t(1-t)} \big[  \trp_{AB} \,  \rho_{ABC}^{1-t}  \, \gamma_{AB}^t  \, - 
                          \trp_B \,   \rho_{BC}^{1-t}  \ \, \gamma_{B}^t \big]  ~ \geq  ~ 0    
                                            \ee
                   where it is important to retain the factor $ \tfrac{1}{t(1-t)}$ which changes sign at $t = 0, 1$.
                   When $\hil_B$ is one-dimensional  \eqref{WYD} implies
              \be
                           \tfrac{1}{t(1-t)} \trp_{A} \,  \rho_{AC}^{1-t} \,  \gamma_{A}^t  & \geq &   \tfrac{1}{t(1-t)}   \,    \rho_{C}^{1-t}     
                                     \ee
                              
                \subsection{But Cauchy-Schwarz matrix inequalities are not new}   \label{sub:CS}
                
                Using $g(x) = (x-1)^2 $ is equivalent to using $x^2$ since the linear terms cancel.
                This gives
                \be
                    \trp_{AB}  \, P_{AB}^2 Q_{ABC}^{-1}  -  \trp_B \, P_B^2 Q_{BC}^{-1}   ~ \geq ~ 0
              \ee
              Since $P_{AB} $ does not depend upon $\hil_C$ one can use the cyclicity of 
              the trace to rewrite this in a more symmetric form as
                 \be   \label{newCS} 
                    \trp_{AB}  \, P_{AB} \, Q_{ABC}^{-1} \, P_{AB} \geq  \trp_B \, P_B  \, Q_{BC}^{-1} \, P_B
              \ee
            However, the inequality \eqref{newCS} is {\em not} new; indeed when $\hil_B$ is 
            one-dimensional it reduces to something slightly less general than
               \be   \label{CS}
                   \trp_A \, X_{AC}^* \, Q_{AC}^{-1} \, X_{AC} ~ \geq ~  X_C^*\,  Q_C^{-1} \, X_{C}
               \ee
                   which was  proved\footnote{In \cite{LbR2} Lieb and Ruskai proved the slightly more
                   general result that $[\Phi(X)]^* {\Phi(A)}^{-1} \Phi(X) \leq  \Phi\big( X^* A^{-1} X \big) $
                   for a completely positive map $\Phi$, of which the partial trace is a special case.   Later, Choi  \cite{Choi}
                   showed that the hypothesis could be weakened to 2-positivity.   The special case
                   $\Phi(X)^* \Phi(X) \leq \Phi(X^* X)$ was shown earlier for unital maps by Kadison and
                   played an important role in Petz's work \cite{Pz1,OP,HMPB}.}
                   in \cite{LbR2}   with  $X-{AC}$ arbitrary 
                   and  $Q_{AC} $ positive 
                   semi-definite with $\ker Q_{AC} \subseteq  \ker X_{AC}^*$.
                   Moreover, this is equivalent to the joint operator
                     convexity of the map $(X,P) \mapsto  X^* P^{-1} X $ also proved in \cite{LbR2}
                     by Lieb and Ruskai, who were unaware until  2010 that the latter had been proved much earlier by
                     Kiefer \cite{Kf}  in 1957.   
                     
                     However, the slightly modified joint convexity
                          \be   \label{xpq}
                        (X,P,Q) \mapsto \tr  X^* \frac{1}{L_P +  t R_Q} X   \qquad  \forall ~ t \in (0, \infty)
                    \ee
                     does not hold as an operator inequality (even for $P = Q$) without the trace.
                    Both \eqref{CS} and \eqref{xpq} can be proved by   very elementary and similar
                    arguments, 
                    as shown  in  \cite{LbR2}  for the former and  for the latter in  \cite{simpSSA} and  the Appendix of \cite{JR}.
                   One can use \eqref{xpq} to prove subadditivity and related inequalities, but the 
                    operator inequality \eqref{CS} does not suffice.   This subtle difference makes 
                  it even more surprising  that Kim's method allows one to essentially extract   operator inequalities from
                  \eqref{xpq} in the bi-partite ans tri-partite settings.

               \subsection{More examples of new inequalities}  \label{sub:more}
                    
                    Although the functions mentioned above are the most commonly considered, there are many more.  
                    As shown in \cite{LesR}, any operator convex  function $k(x): (0, \infty ) \mapsto (0,\infty) $ satisfying 
                    the symmetry condition $x k(x) = k( x^{-1})$ defines an operator convex function $g(x) = (1 - x)^2 k(x) $ 
                    with the symmetry property
                    $\wtd{g}(x) = x \, g(x^{-1}) = g(x) $ which can be used to define  an $H_g(K,P,Q) $
                    as above.   The symmetrization  
                    $$g(x) + \wtd{g}(x) = - \log x + x \log x = (x-1) \log x $$
                     yields  $k(x) = (\log x)/(x-1)$.   However, the symmetrized version  yields a less
                     transparent inequality  since one would have the sum of \eqref{SSAposA}
                     and \eqref{SSArev}.
                                         
                                         Several families of   functions  $k$ have been studied by Petz \cite{Pz2} (who uses $f = 1/k$
                                          operator monotone)  in the context of
                    monotone Riemmanian metrics and, more recently, by Hiai and Kosaki \cite{HK1,HK2} who developed a theory of
                    operator means.  A fairly comprehensive  list is given in \cite[Section~4]{HKPR}.  
                    
                      The symmetrized version of the primitive example in Section~\ref{sub:CS} is  
                      $$
                        \half[ g(x) + \wtd{g}(x)]  = (1-x)^2  \frac{1+x}{2x}
                        $$ 
                        which yields $k(x) = (1 + x)/2x$.
                     It is well-known \cite{Pz2,LesR,HKPR} that the functions $k(x) $ satisfy a partial order with 
                    \be
                     \frac{2}{1+x} \leq k(x) \leq \frac{1+x}{2x}
                     \ee
                    The smallest element $k(x) = 2/(1+x) $ is associated with the Bures
                    metric  but 
                    $g(L_P R_Q^{-1}) R_Q   = (L_P - R_Q)^2(/L_P + R_Q)$ does not seem to yield
                     particularly transparent inequalities when inserted in \eqref{final}.   
                  
                         The function
                     $k(x) = x^{-1/2} $ also plays a special role in some situations \cite{HK1,HK2, HKPR,TKRWV}.
      In this case, we can
                                          ``unsymmetrize'' to $g(x) =  x^{-1/2} - x^{1/2}$    and $\wtd{g}(x) = x ( x^{1/2} - x^{-1/2}  )  $   
                   to obtain the inequality
                \be   \label{xhalf}
                       \trp_{AB} \,  \gamma_{AB}^{-1/2} [ \rho_{ABC} - \gamma_{AB} ] \rho_{ABC}^{1/2}    
                      -   \trp_{B}  \, \gamma_{B}^{-1/2} [ \rho_{BC} - \gamma_{B} ] \rho_{BC}^{1/2} ~ \geq ~ 0                   
                             \ee
                     where we used $g$ with $P_{AB} = \gamma_{AB}, ~ Q_{ABC} = \rho_{ABC} $.
                    
                                                             
                       \section{Equality conditions}
                       It is natural to ask under what conditions equality holds in these inequalities.  In the case
                       of those related to SSA, i.e.,  \eqref{SSApos} and \eqref{SSArev}, it is easy to see that the equality conditions
                       given in \cite{HJPW} suffice.   In the simplest case,  $\hil_B = \hil_{B^\prime} \ot \hil_{B^{\prime \prime} } $
                       and $\rho_{ABC} = \rho_{AB^\prime} \ot \rho_{B^{\prime \prime } C} $.   The general case is a direct 
                       sum of this situation, i.e.,  $\hil_B =  \bigoplus_k \hil_{B^\prime}^k \ot  \hil_{B^{\prime \prime}}^k  $ and
                       \be   \label{eqcond}
                          \rho_{ABC} =    \bigoplus_k  \rho_{AB^\prime}^k \ot \rho_{B^{\prime \prime } C}^k
                       \ee
                       Since a positive semi-definite matrix  $A \geq 0$ is  equal to zero if and only if $\tr A = 0$,
                       it is immediate that \eqref{eqcond} is necessary and sufficient for equality.
                       This also gives    conditions for equality in \eqref{final} for
                      the other examples  with  $Q_{ABC} = \rho_{ABC} $ and $  P_{AB} = \rho_{AB} $ 
                      and are essentially
                       independent of the function $g$.
                       This is because  Nevanlinna's theorem  \cite[Section 59, Theorem 2]{AG}
                        implies that any  operator convex function $g$ on $(0,\infty)$ with $g(1) = 0 $
                       has an  integral representation of the general form
                       \be   \label{intrep}
                            g(x) = a x + b x^2 + \int_0^\infty  \frac{ f(x,t) }{x + t}    ~   d\mu_g(t)
                       \ee
                       (The precise representations are written  in equivalent, but slightly differently  forms
                       in several references, including eq. (8.2)  in \cite{HMPB} or eq. (17) in  \cite{JR}
                        or eq. (13) in \cite{LesR}.   The details are not relevant here.) 
                        Whenever the corresponding measure $\mu_g(t)$ is supported on $(0,\infty)$, the conditions 
                      \eqref{eqcond} are necessary and sufficient for equality.   
                           In the approach of \cite{JR}  the equality conditions arise
                       as the condition for
                      equality in the joint convexity in \eqref{xpq}   for all   $t \in (0, \infty)$, which
                      makes the somewhat surprising lack of dependence on $g$ transparent.  
                      
                      For inequalities with more general choices of $P_{AB}$ in  \eqref{final}, 
                      as in Sections ~\ref{sub:MPT}, \ref{sub:WYD} and
                      \ref{sub:more},   $\gamma_{AB} $ must also have a similar block representation
                    with  $ \rho_{AB^\prime}^k  = \gamma_{AB^\prime}^k  $ for equality.                    

               \section{Historical remarks}
                    
                    The proofs of joint convexity in $P,Q$ of functions of
                    the type defined in \eqref{Hdef} for
                    operator convex functions $g \in \cG$ are based on the relative modular 
                    operator  $\Delta_{PQ} = L_P R_Q^{-1} $ introduced by Araki in
                    a much more general context.   
                    The operator $L/R$   in Effros's perspective   is essentially   $\Delta_{PQ} $.    
                    Using an integral representation  of the form \eqref{intrep}
                     one can reduce the joint convexity of $(P,Q) \mapsto H_g(K,P,Q)$ to
                    the joint convexity of  the map in \eqref{xpq} which, as mentioned above,
       can be proved by a very elementary argument, as shown in
                    \cite{simpSSA} and  the Appendix of \cite{JR}.
                    
                    Uhlmann \cite{U} seems to have been the first to realize that one 
                    could take the partial trace by integrating unitary conjugations over
                    Haar measure.   A pedestrian equivalent is to use the discrete 
                    Weyl-Heisenberg group (as in, e.g., \cite{JR}),  a process sometimes called
                    ``twirling'',  although
                    other orthogonal unitary bases can also be used \cite{Ty}.  
                     Remarkably, Uhlmann  \cite{U} also realized that the concavity of
                    the map $A \mapsto   \tr  e^{K + \log A} $ could then be used to prove
                    SSA.   Without knowing about Uhlmann's work, Lieb found and proved this concavity  
                    \cite[Theorem~6 ]{Lb} which was the key
                    ingredient in the two original proofs of SSA presented in \cite{SSA},
                    both of which are different from Uhlmann's.  
                                        
                    Instead of a two-step argument using joint convexity and unitary
                    conjugation, one can go directly to the monotonicity under quantum
                    channels, i.e., completely positive trace-preserving maps of which
                   the partial trace is a special case.   This was first done by Petz under the
                   slightly weaker condition\footnote{To be more precise, Petz uses the 
                   weak Kadison form of the operator Schwarz inequality in Section~\ref{sub:CS} which follows 
                   from  2-positivity.  See \cite{HMPB} for details.}  of 2-positivity
                   and a form of Jensen's inequality
                    \cite{ HMPB,OP,Pz1}.  Another argument was given in \cite{LesR} based
                    on the integral representation of convex operator functions 
                    and an elementary Schwarz argument
                     similar to the  one in \cite{simpSSA} and \cite[Appendix]{JR}.
                  
                     One advantage of this approach is that one does not need to 
                     add the 
                    $ \trp_A  \,  \tfrac{1}{d_A} $ to the RHS of \eqref{monoa} as   Kim did in  \cite{kim}  to perform
                     twirling.  (See also \cite{JR}.)
                     One can prove \eqref{monoa} directly
                and then use  
                $ \tr_{AB}  \,   (\quad)_{AB}   \, Q_{AB}    =   \tr_{ABC}  \, (\quad )_A  \,Q_{ABC} $
              as in the last line of \eqref{mono}.
       
                     
                     It is also well-known that one can use a block matrix representation
                     to show that monotonicity under partial traces implies convexity, as
                     noted in \cite{Lb,SSA,MBR,W}.   See \cite{MBR,W} for additional
                     background.
                                         
                         \bigskip
                     
                     \noindent{\bf Acknowledgment:}  It is a pleasure to thank Jon Tyson
                     for drawing my attention to \cite{kim} and for subsequent stimulating discussions.
                     This work was partially supported by NSF Grant CFF-1018401  which is
                  administered by Tufts University.  
                    

\end{document}